\documentclass[12pt]{article}
\usepackage{epsf}

\begin{document}

\title{\LARGE \bf Quasi-SU(3) truncation scheme \\ for even-even $sd$-shell
nuclei}

\author{
C. E. Vargas$^1$\thanks{Electronic address: cvargas@fis.cinvestav.mx},
J. G. Hirsch$^2$\thanks{Electronic address: hirsch@nuclecu.unam.mx},
J. P. Draayer$^3$\thanks{Electronic address: draayer@lsu.edu},\\
{\small $^1$  Departamento de F\'{\i}sica, Centro de Investigaci\'on y de
Estudios Avanzados del IPN,}\\
{\small Apartado Postal 14-740 M\'exico 07000 DF, M\'exico}\\
{\small $^2$  Instituto de Ciencias Nucleares, Universidad Nacional
Aut\'onoma de M\'exico,}\\
{\small Apartado Postal 70-543 M\'exico 04510 DF, M\'exico }\\
{\small $^3$  Department of Physics and Astronomy, Louisiana State
University,}\\
{\small Baton Rouge, LA 70803-4001, USA }
}

\date{\today}

\maketitle

\begin{abstract}
{\bf Abstract} The quasi-SU(3) symmetry was uncovered in full {\em pf} and
{\em sdg} shell-model calculations for both even-even \cite{Zuk95} and
odd-even nuclei \cite{Mar97}. It manifests itself through a dominance of
single-particle and quadrupole-quadrupole terms in a Hamiltonian used to
describe well-deformed nuclei. A practical consequence of the quasi-SU(3)
symmetry is an efficient basis truncation scheme. In \cite{Var98} it is shown
that when this type of Hamiltonian is diagonalized in an SU(3) basis, only a
few irreducible representations (irreps) of SU(3) are needed to describe the
yrast band, the leading S=0 irrep augmented with the leading S=1 irreps in
the proton and neutron subspaces.
  In the present article the quasi-SU(3) truncation scheme is used, in
conjunction with a ``realistic but schematic'' Hamiltonian that includes the
most important multipole terms, to describe the energy spectra and B(E2)
transition strengths of $^{20,22}$Ne, $^{24}$Mg and $^{28}$Si. The effect of
the size of the Hilbert space on both sets of observables is discussed, as
well as the structure of the yrast band and the importance of the various
terms in the Hamiltonian.

\end{abstract}

\vskip2pc

\section{Introduction}

More than 40 years ago Elliott pointed out the fundamental role SU(3) plays
in a description of light rotational nuclei \cite{Ell58}. The introduction
of SU(3) yields insight into relevant degrees of freedom underlying collective
rotational motion, and in particular it points to the importance of the
quadrupole-quadrupole interaction. Furthermore, even though the spin-orbit
interaction breaks SU(3), it can still be used to truncate full model spaces
into small subspaces in which realistic calculations can be performed.

In heavy deformed nuclei the pseudo-SU(3) symmetry \cite{Ari69} has played a
similar role. While in the 1980s it was employed as an exact symmetry
\cite{Dra82,Dra84,Cas87,Cas88}, after the development of new computer
codes for calculating reduced matrix elements of 1- and 2-body operators
\cite{Bah94}, it was possible to take into account pseudo-SU(3) symmetry
breaking terms in the Hamiltonian (i.e. realistic single-particle energies
and the pairing interaction) which have been shown to be important in a
description of low-energy bands and associated electromagnetic transition
strengths in even-even \cite{Rom98,Beu98,Beu00} as well as odd-mass
\cite{Var99,Var00} nuclei.

Although the pseudo-SU(3) model is a successful shell-model scheme, it is
limited by the formal exclusion of valence nucleons occupying unique parity
orbitals. In certain cases a reparametrization of the theory can be used to
compensate for this exclusion, while in other cases the nucleons in unique
parity orbitals have to be included explicitly \cite{Esc95}. The quasi-SU(3)
truncation scheme therefore offers the possibility of including nucleons in
intruder orbits within the framework of a complementary SU(3) formalism and
thereby yielding a complete theory for heavy deformed nuclei \cite{Zuk95}.

The quasi-SU(3) symmetry was uncovered in full {\em pf} and {\em sdg}
shell-model calculations for even-even \cite{Zuk95} and odd-even \cite{Mar97}
nuclei. It owes its importance to the dominance of the single-particle and
quadrupole-quadrupole terms in a Hamiltonian used to describe well-deformed
nuclei. Needless to say, other terms such as pairing are crucial in
determining observed moments of inertia, but most of these effects can be
accounted for perturbatively because they introduce small changes in the
wavefunctions \cite{Cau95}.

The quasi-SU(3) symmetry also leads to an efficient truncation scheme. In
shell-model calculations it has been shown that the single-particle levels
with $j = l + {\frac 1 2}$ play a dominant role in the low-energy spectra,
allowing significant reductions in the size of the Hilbert space
\cite{Zuk95}.  In \cite{Var98} it is reported that, when a single-particle
plus quadrupole-quadrupole Hamiltonian is diagonalized in an SU(3) basis,
very few SU(3) irreducible representations (irreps) are needed to describe
the yrast band. The ground state band is built from the S=0 leading irrep,
which strongly mixes with the leading S=1 irreps in the proton and neutron
subspaces. Using a realistic Hamiltonian the change in the wavefunction
associated with backbending in $^{48}$Cr was studied in \cite{Hir99}.

In the present article the quasi-SU(3) truncation scheme obtained in
\cite{Var98} is used in conjunction with a realistic Hamiltonian.
The Hamiltonian has single-particle terms for protons and neutrons,
quadrupole-quadrupole and pairing interactions, and three rotor-like terms
that allow for a fine tuning of the different bands. The energy spectra and
B(E2) transition strength of $^{20,22}$Ne, $^{24}$Mg and $^{28}$Si are
studied. The effects of the size of the Hilbert space on both observables
are discussed at length, as well as the yrast band wavefunction and the role
played by the different terms in the Hamiltonian.

The main goal is to show that the combination of the quasi-SU(3) truncation
scheme and a realistic but simple Hamiltonian is a powerful combination for
generating a description of the low-energy properties of deformed nuclei.
This work, and a follow-on one where odd-mass and odd-odd 
light nuclei are discussed
\cite{Var00b}, provide justification for applications of the quasi-SU(3)
truncation scheme to nucleons occupying unique parity orbitals in heavy
deformed nuclei.

We study two nuclei, for which a diagonalization in the full Hilbert space in
the SU(3) scheme is feasible ($^{20,22}$Ne), to analyse the effects of the
truncation on the energy spectra and B(E2) values. The heavier $^{24}${Mg}
and $^{28}${Si} nuclei have a richer structure and are used to help exhibit
the efficiency of the model. The last two nuclei are more affected by the
spin-orbit interaction, which introduces significant mixing of SU(3) irreps
in the yrast band wavefunction. We will also discuss changes in the rotor-like
terms in the Hamiltonian which allow for a best description of the nuclear
properties in each truncated subspace.

The article is organized as follows: In Section 2 a brief review of the
SU(3) basis and the model Hamiltonian is presented. Results for $^{20}$Ne
are shown and discussed in Section 3, for $^{22}$Ne in Section 4, for
$^{24}$Mg in Section 5, and for $^{28}$Si in Section 6. Conclusions are
given in Section 7.

\section{The quasi SU(3) model}

There are many articles and books were the SU(3) nuclear model is discussed
extensively and to which the interested reader is referred
\cite{Ell58,Mosh67,Cas93,Bah95,Tro95,Tro96}. In the present section we
review some properties of the model which are of concern in the present work.

The basis states are written as
\begin{eqnarray}
| \{ n_\pi [f_\pi] \alpha_\pi (\lambda_\pi,\mu_\pi), n_\nu [f_\nu] \alpha_\nu
(\lambda_\nu,\mu_\nu) \}
\rho (\lambda,\mu) k L \{S_\pi,S_\nu\} S ; JM \rangle \label{states}
\end{eqnarray}
where $n_\pi $ is the number of valence protons in the {\em sd} shell and
$[f_\pi] $ is the irrep of the U(2) spin group for protons, which has
associated with spin $\ S_\pi = (f^1_\pi - f^2_\pi)/2$. The SU(3) irrep
for protons is $(\lambda_\pi,\mu_\pi)$ with a multiplicity label $\alpha_\pi$
associated with the reduction from $U(6)$. Similar definitions hold for the
neutrons, labeled with $\nu$. There are other two multiplicity labels:
$\rho$, which counts how many times the total irrep $(\lambda, \mu)$ occurs in
the direct product $(\lambda_{\pi}, \mu_{\pi}) \otimes (\lambda_{\nu},
\mu_{\nu})$ and $K$, which classifies the different occurrences of the
orbital angular momentum
$L$ in $ (\lambda, \mu)$.

The vector states (\ref{states}) span the complete shell-model space within
only one active (harmonic oscillator) shell for each kind of nucleon. As an
example, for two protons ($n_\pi = 2$) in the {\em sd} shell there are three
possible irreps: $(\lambda_{\pi}, \mu_{\pi})=$ (4,0), (2,1) and (0,2).
The first and third irreps have spin zero, the second one has spin 1.
Each one occurs only once ($\alpha_\pi = 1$). The same numbers are
obtained for 2 neutrons. The coupled SU(3) irreps are ordered by decreasing
values of the expectation value of the second order Casimir operator, $C_2$,
\begin{equation}
\langle (\lambda,\mu) | C_2 | (\lambda,\mu) \rangle
= (\lambda +  \mu +3) ~ (\lambda + \mu) - \lambda \mu  .
\end{equation}
They are listed in Table 1, with the $(\lambda,\mu)$ labeling the coupled
irreps in the third column, the possible values of the total spin S in
the fourth, and the $C_2$ value in the fifth column.

\begin{table}
\begin{tabular}{ccccc|ccccc}
$(\lambda_\pi, \mu_\pi )$&$(\lambda_\nu, \mu_\nu )$&$(\lambda, \mu )$&
$S$ & $C_2$
&$(\lambda_\pi, \mu_\pi )$&$(\lambda_\nu, \mu_\nu )$&$(\lambda, \mu )$&
$S$ & $C_2$
  \\ \hline
(4,0) & (4,0) & (8,0) & 0   & 88       &
(4,0) & (4,0) & (6,1) & 0   & 64       \\
(4,0) & (2,1) & (6,1) & 1   & 64       &
(2,1) & (4,0) & (6,1) & 1   & 64       \\
(4,0) & (4,0) & (4,2) & 0 & 46       &
(4,0) & (2,1) & (4,2) & 1 & 46       \\
(2,1) & (4,0) & (4,2) & 1 & 46       &
(4,0) & (0,2) & (4,2) & 0 & 46       \\
(0,2) & (4,0) & (4,2) & 0 & 46       &
(2,1) & (2,1) & (4,2) & 0,1,2 & 46       \\
(4,0) & (2,1) & (5,0) & 1 & 40       &
(2,1) & (4,0) & (5,0) & 1 & 40       \\
(2,1) & (2,1) & (5,0) & 0,1,2 & 40    &
(4,0) & (4,0) & (2,3) & 0 & 34      \\
(4,0) & (2,1) & (2,3) & 1 & 34       &
(2,1) & (4,0) & (2,3) & 1 & 34       \\
(2,1) & (2,1) & (2,3) & 0,1,2 & 34      &
(2,1) & (0,2) & (2,3) & 1 & 34      \\
(0,2) & (2,1) & (2,3) & 1 & 34       &
(4,0) & (4,0) & (0,4) & 0 & 28      \\
(2,1) & (2,1) & (0,4) & 0,1,2 & 28     &
(0,2) & (0,2) & (0,4) & 0 & 28     \\
(4,0) & (2,1) & (3,1) & 1 & 25     &
(2,1) & (4,0) & (3,1) & 1 & 25     \\
(4,0) & (0,2) & (3,1) & 0 & 25     &
(0,2) & (4,0) & (3,1) & 0 & 25     \\
(2,1) & (2,1) & (3,1) & 0,1,2 & 25   &
(2,1) & (0,2) & (3,1) & 1 & 25    \\
(0,2) & (2,1) & (3,1) & 1 & 25     &
(4,0) & (2,1) & (1,2) & 1 & 16    \\
(2,1) & (4,0) & (1,2) & 1 & 16    &
(2,1) & (2,1) & (1,2) & 0,1,2 & 16     \\
(2,1) & (0,2) & (1,2) & 1 & 16    &
(0,2) & (2,1) & (1,2) & 1 & 16     \\
(0,2) & (0,2) & (1,2) & 0 & 16    &
(4,0) & (0,2) & (2,0) & 0 & 10    \\
(0,2) & (4,0) & (2,0) & 0 & 10     &
(2,1) & (2,1) & (2,0) & 0,1,2 & 10     \\
(2,1) & (0,2) & (2,0) & 1 & 10    &
(0,2) & (2,1) & (2,0) & 1 & 10     \\
(0,2) & (0,2) & (2,0) & 0 & 10     &
(2,1) & (2,1) & (0,1) & 0,1,2 &  4    \\
(2,1) & (0,2) & (0,1) & 1 &  4   &
(0,2) & (2,1) & (0,1) & 1 &  4   \\ \hline
\label{eq:irrepsne20}
\end{tabular}
\caption{Complete list of irreps for $^{20}{Ne}$. The proton
$(\lambda_\pi, \mu_\pi )$, neutron $(\lambda_\nu, \mu_\nu )$ and coupled irreps
$(\lambda,\mu)$ are listed in the first three columns, the values of
the total spin S in the fourth, and the $C_2$ value in the fifth column.}
\end{table}
There is a total of 66 irreps, including the outer multiplicity $\rho=1,2$
in the couplings $(2,1)_\pi \otimes (2,1)_\nu = (3,1), (1,2)$,
and the total spin S. For each irrep and each spin there are many
states, labeled by their orbital angular momentum $L$, its multiplicity
$K$, their total angular momentum $J$ and its projection $M$.


For $^{20}${Ne} and $^{22}${Ne} all possible states were included. For
$^{24}${Mg} and $^{28}${Si} the space was truncated to allow for faster
calculations and to analyze the validity of a truncation scheme based on
the quasi-SU(3) basis. The quasi SU(3) truncation scheme is very simple
\cite{Var98}: only the SU(3) irreps with the largest $\langle C_2 \rangle$
value, in the separate proton and neutron spaces as well as the coupled space,
and spin 0 and 1 are included. The inclusion of spin 1 states represents 
the most important difference from previous SU(3) based truncation schemes 
\cite{Tro95}.

The Hamiltonian of the model is
\begin{eqnarray}
  H & = & H_{sp,\pi} + H_{sp,\nu} - \frac{1}{2}~ \chi~ Q \cdot
          Q - ~ G_\pi ~H_{pair,\pi} ~\label{eq:ham} \\
    &   & - ~G_\nu ~H_{pair,\nu} + ~a~ K_J^2~ +~ b~ J^2~ +~ A_{sym}~
          \tilde C_2 . \nonumber 
\end{eqnarray}

\noindent The first terms are spherical Nilsson single-particle energies,
the quadrupole-quadrupole  and pairing interactions.
They are the basic components of any realistic Hamiltonian \cite{Rin79,Duf96}
and have been widely studied in the nuclear physics literature, allowing
their respective strengths to be fixed by systematics \cite{Rin79,Duf96}.
The remaining are three rotor-like terms used to fine tune the moment of
inertia and the position of the different $K$ bands. The SU(3) mixing is due to
the single-particle and pairing terms.

The single-particle part of the Hamiltonian is
\begin{equation}
H_{sp} = \hbar \omega_0 \left( \eta + \frac{3}{2} -
2 \kappa  \, \vec{l} \cdot \vec{s} + \kappa \mu \, \vec{l}^2 \right),
\label{Nilssonh} \end{equation}
with parameters \cite{Rin79}
\begin{eqnarray}
\hbar {\omega}_0 = 41 A^{-1/3} [MeV],
&\kappa_\pi =  \kappa_\nu = 0.08,
&\mu_\pi = \mu_\nu = 0.0 .
\end{eqnarray}
\noindent The pairing interaction is
\begin{equation}
V_p = -\frac{1}{4} G \sum_{j,j'} a^\dagger_j
a^\dagger_{\bar{j}} a_{j'} a_{\bar{j'}} \label{pair}
\end{equation}
where $\bar{j}$ denotes the time reversed partner of the single-particle state
$j$ and $G$ is the strength of the pairing force. Its second 
quantized expression
in term of SU(3) tensors is reviewed in \cite{Tro95,Var00}.
\noindent The quadrupole-quadrupole ($\chi$) and pairing ($G_{\pi,\nu}$)
interaction strengths used are \cite{Rin79,Duf96}
\begin{equation}
\chi = \frac{17}{A^{5/3}},~~~~~~~~G_{\pi} = G_{\nu} = \frac{9.5}{A} .
\end{equation}
The pairing force parameter (9.5) in these light nuclei ($A \approx 17 -
28$) is about half the one used in heavier nuclei.

The three `rotor-like' terms have been studied in detail in previous papers
where the SU(3) and pseudo SU(3) symmetry were used as a dynamical symmetries
\cite{Dra84,Cas87}. In the present work, $a$ and $b$ are the two parameters
used to fit the spectra. $A_{sym}$ was used for $^{22}$Ne and $^{24}$Mg. The term
proportional to $K_J^2$ breaks the SU(3) degeneracy of the different K bands
\cite{Naq90}. The term proportional to $J^2$ is used to fine tune the moment
of inertia. It represents a small correction to the  quadrupole-quadrupole
term, which contributes to the rotor spectra with strength $3/2 \chi$
\cite{Var00}. The symmetry term distinguishes SU(3) irreps with both
$\lambda$ and $\mu$ even from the others \cite{Les87}, having zero strength
in the first case and a positive value in the the second. The net effect of
this term is to make contributions of irreps with both $\lambda$ and $\mu$
even enhanced relative to the other because they belong to different symmetry
types of the intrinsic Vierergruppe $D_2$ \cite{Les87}.

The values of the Hamiltonian parameters used for the different nuclei
are listed in Table 2. The $a, b$ and $A_{sym}$ paramenters were adjusted to
obtain a best fit for each Hilbert space, which is indicated as {\em full}
or {\em truncated}. In $^{28}$Si we also studied a Hamiltonian with an extra
term proportional to the third order Casimir $C_3$ (see Section 6).

\begin{table}
\begin{tabular}{cc|ccccc}
    nuclei &   Space   & $\chi$ &$G_{\pi,\nu}$ & $A_{sym}$ & $a$  & $b$
\\ \hline
$^{20}$Ne & full      & 0.1154  &  0.4750      &     0      & 0.6  & -0.010 \\
$^{20}$Ne & truncated & 0.1154  &  0.4750      &     0      & 0.3  &    0   \\
$^{22}$Ne & full      & 0.0984  &  0.4318      &     0      & 0    & -0.030 \\
$^{22}$Ne & truncated & 0.0984  &  0.4318      &   0.010    & 0.20 & -0.020 \\
$^{24}$Mg & truncated & 0.0851  &  0.3958      &   0.08     & 0.56 &  0.023 \\
$^{28}$Si & truncated & 0.0658  &  0.3393      &     0      & 0.30 & -0.036
\label{t1}
\end{tabular}
\caption{Parameters of Hamiltonian (\ref{eq:ham}) for the different nuclei
listed in the first column. Labels ``full'' or ``truncated'' refer to the
Hilbert space used. }
\end{table}

In the following sections we present results for the nuclei $^{20}$Ne,
$^{22}$Ne, $^{24}$Mg and $^{28}$Si. The effects of both the truncation of
the Hilbert space and the rotor-like terms of the Hamiltonian (\ref{eq:ham})
on the energy spectra and the B(E2) electromagnetic transition strengths
are investigated. When possible, a comparison with experimental data
\cite{NNDC} and full sd-shell model calculations \cite{Pre72,Wil84} is given.

\section{The $^{20}$Ne case}

$^{20}$Ne is considered to be a nucleus with 2 protons and 2 neutrons in the
sd-shell, outside an inert $^{16}$O core. The list of the 66 SU(3) irreps
that define the full Hilbert space is shown in Table 1.

The energy spectra of $^{20}$Ne obtained with Hamiltonian (\ref{eq:ham})
diagonalized in this basis is shown in the second column of Fig. 1, labeled
`Full'. A truncated Hilbert space built with only the 12 irreps with the
largest $C_2$ values (listed in the upper six lines in Table 1) was also
used. The energy spectra is shown in the first column of Fig. 1, labeled
`Trunc'. For the full space the rotor parameters were $A_{sym} = 0, b =
-0.01, a = 0.6$ MeV and for the truncated space $A_{sym} = b = 0, a = 0.3$
MeV. The theoretical results are compared with
the experimental energies \cite{NNDC}, shown in the third column,  and
with the energies obtained in previous theoretical studies using an SU(3)
basis restricted to S=0 states \cite{Tro96} and the full sd shell-model
\cite{Wil84}. In the figure, a) includes the ground state as well as the
$\beta$ and $\gamma$ bands, while b) shows other excited configurations.

\begin{figure}   
\epsfxsize= 18cm
\centerline{\epsfbox{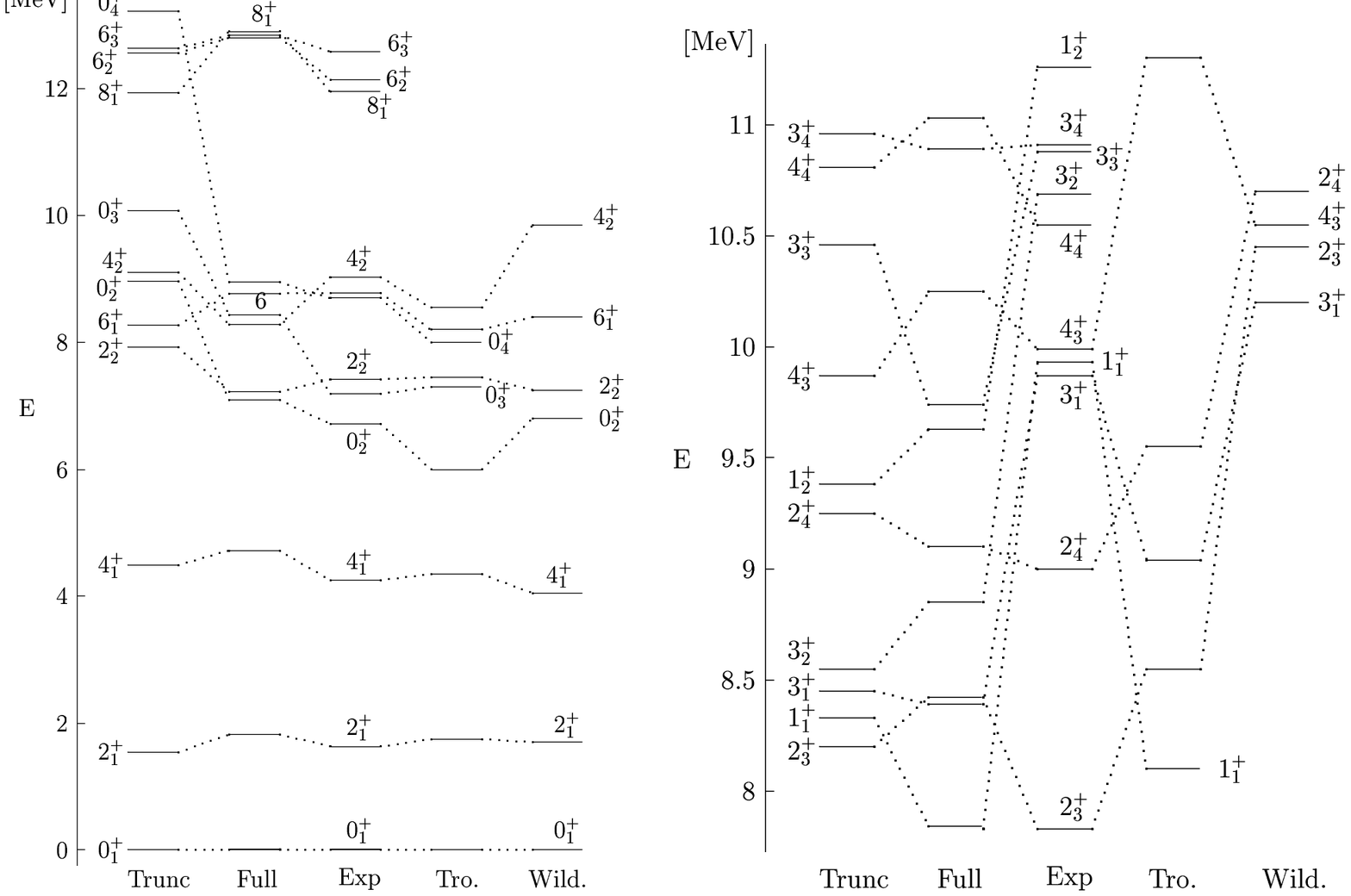}}\vspace*{-11.6cm}
\caption{ Energy spectra of $^{20}$Ne. For details see the text. The label
Tro refers to  Ref.~\cite{Tro96} and Wild to Ref.~\cite{Wil84}. Note that
there is a change in scale between a) to b).}
\label{fig1}
\end{figure}

The ground state band ($0^+_1, 2^+_1, 4^+_1, 6^+_1, 8^+_1$) is described
almost perfectly by all the models. The $\gamma$ band, which starts with the
$0^+_2$, and the $\beta$ band, which starts with the $2^+_2$ state, are
also described well, however, in the truncated basis their order is reverted.
The excited $4^+_2, 6^+_2, 6^+_3$ are also reproduced well.

The excited $0^+_3, 0^+_4$ states have energies around 8 MeV. The full basis
yields a correct description, while with the truncated space their energies
are too high. This is a clear indication of a limitation of the truncated
space, namely, it cannot be used to describe highly excited states.

Part b) displays many excited states. While some of them are correctly
reproduced, as for example $2^+_4, 4^+_3, 4^+_4$, other are not. The
$1^+$ states are some of the worst described. The energies reported in
\cite{Tro96}, fourth column, are less accurate in their description of
some excited states due to the absence of $S=1,2$ states in the basis.
It is important to note that even the full space calculation shown in
the fifth column does not give a good description of these states.

To analyze the effect the truncation of the Hilbert space has on the
B(E2) transition strengths, in Table 3 B(E2) values are shown as a function
of the number of SU(3) irreps included in the basis. These numbers (1, 4, 12,
17, 25, 66) were selected based on changes in the values of $C_2$ in
Table 1, guaranteeing in this way that irreps which contribute with the
same intensity to the quadrupole-quadrupole force are all included or
all neglected. The first column shows the initial and final angular
momentum state of the transition, the second the experimental B(E2) value,
and the remaining six the theoretical B(E2) values obtained using the same
Hamiltonian parameters but reducing the size of the basis (indicated at the
top of each column). The last column, labeled 12$^*$, shows the B(E2) values
obtained using a space with 12 irreps (as for the sixth column) but the
Hamiltonian parameters optimized for this space, namely, the same set used to
calculate the energy spectra shown in Fig. 1.  In all cases an effective
charge of $q_{ef} = 1.558$ was used.

\begin{table}{\footnotesize
\begin{tabular}{c|ccccccc|c}
  $J \rightarrow (J+2)$ &   Exp.              &  66    &   25   &   17 
&   12   &    4   &    1   & 12$^*$ \\ \hline
  $0_1 \rightarrow 2_1$ & $3.2731 \pm 0.1612$ & 2.2373 & 2.3210 & 
2.3802 & 2.4101 & 2.5410 & 2.7312 & 2.4051 \\
  $2_1 \rightarrow 4_1$ & $1.2896 \pm 0.1172$ & 1.0471 & 1.0302 & 
1.0523 & 1.0743 & 1.1424 & 1.2414 & 1.0624 \\
  $4_1 \rightarrow 6_1$ & $0.9316 \pm 0.1397$ & 0.6594 & 0.6745 & 
0.6745 & 0.6802 & 0.7591 & 0.8475 & 0.6670 \\
  $6_1 \rightarrow 8_1$ & $0.3795 \pm 0.5482$ & 0.3274 & 0.3231 & 
0.3261 & 0.3395 & 0.3931 & 0.4527 & 0.3320 \\
  $0_1 \rightarrow 2_7$ & $0.0031 \pm 0.0008$ & 0.0063 & 0.0353 & 
0.0011 & 0.0067 & 0.0033 & --     & 0.0011 \\
  $0_2 \rightarrow 2_1$ & $0.1177 \pm 0.0145$ & 0.0948 & 0.0492 & 
0.0072 & 0.0074 & 0.0047 & --     & 0.0072 \\
  $0_3 \rightarrow 2_1$ & $0.0099 \pm 0.0019$ & 0.0080 & 0.0075 & 
0.0227 & 0.0257 & 0.0085 & --     & 0.0249 \\
  $2_1 \rightarrow 4_8$ & $0.0006 \pm 0.0003$ & 0.0013 & 0.0001 & 
0.0021 & 0.0002 & 0.0001 & --     & 0.0003
\label{eq:be2ne20}
\end{tabular}~\\~
\caption{ B(E2;J $\rightarrow$ (J+2)) [$e^2b^2 \times10^{-2}$] transition
strengths in $^{20}$Ne for different size model spaces that included the
number of SU(3) irreps indicated on the column header. The last column was
obtained with 12 irreps using Hamiltonian parameters that were optimized for
that 12-irrep space.}}
\end{table}

\begin{figure}
\vspace*{-1cm}\hspace*{-0.5cm}
\epsfxsize=14cm
\centerline{\epsfbox{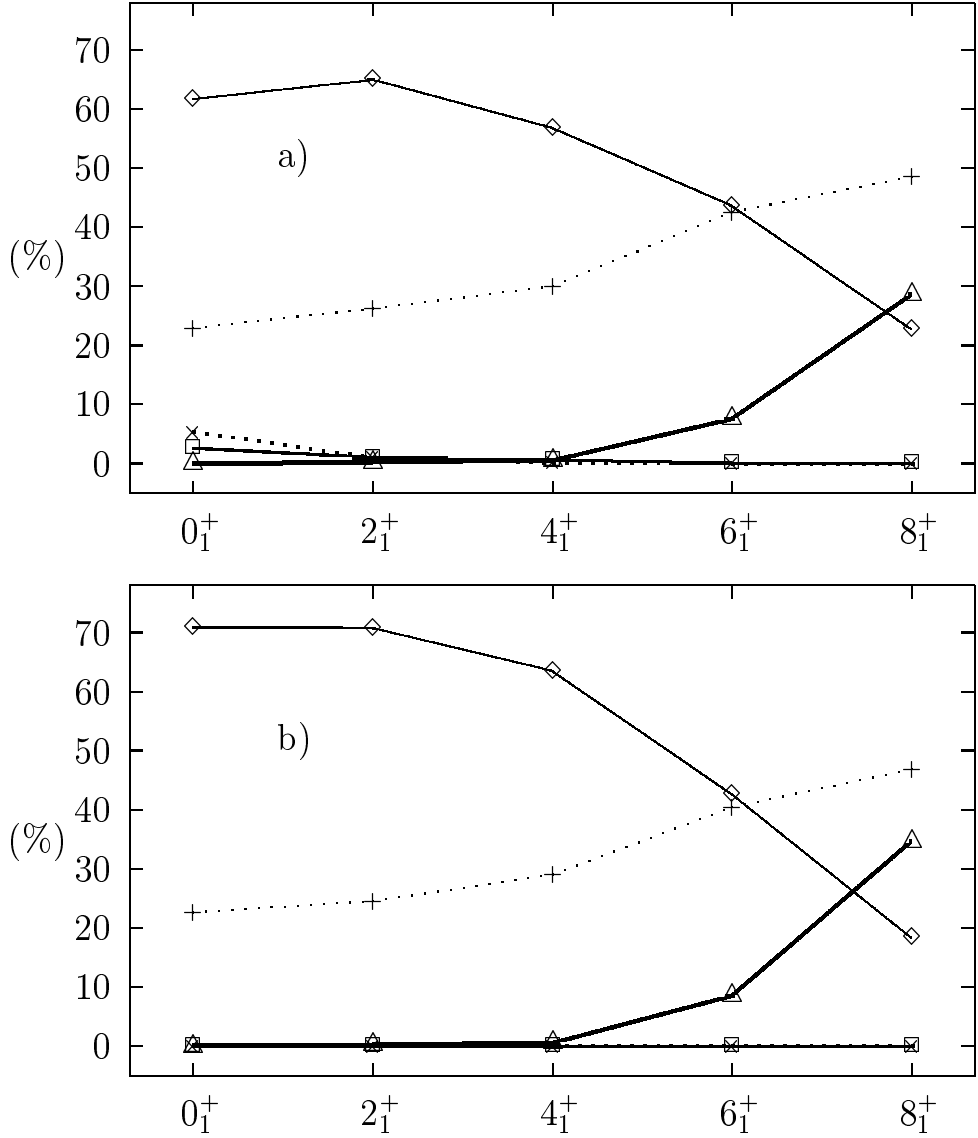}}\vspace*{-3cm}
\caption{Wavefunction components of states belonging to the ground state
band in $^{20}$Ne. The percentage each irrep contributes is shown as a
function of the angular momentum for the complete basis, insert a),  and
the truncated basis, insert b). The convention used is: $\diamond$ for (8,0)0;
$+$ for (6,1)1; $\sqcap \!\!\!\! \sqcup$ for [(4,0)$\otimes$ (4,0)] (4,2)0;
$\times$ for [(4,0)$\otimes$ (0,2)] (4,2)0 and [(0,2)$\otimes$ (4,0)]
(4,2)0; and  $\triangle$ for (4,2)2.}
\end{figure}

For the B(E2) transitions within the ground state band any space produces
very good results. For the same Hamiltonian parameters, the B(E2) values all
increase as the number of irreps in the basis decreases. The results for
12 irreps with the best-fit parameters are as good as those obtained
with the complete basis, and very close to the experimental ones, as can
be seen in the first four rows of Table 3.

For B(E2) transitions from states in excited bands to states in the
ground state band the situation changes drastically. The leading irrep is
(8,0), which has no $\gamma$ band. So including only one SU(3) irrep yields
no excited bands and no transitions. The transition $0_2 \rightarrow
2_1$ is pretty well described with 12 irreps. The $0_3 \rightarrow 2_1$
transition (next to the last entry in Table 3) is is not so well described.
The last entry is only reasonable in the complete basis. For basis including
1 or 4 irreps the $8^+_1$ state always has an energy higher than that
observed because the wavefunction is close to a pure rotor. At least
12 irreps are needed to move this state down to the observed energy.

The percentage each SU(3) irrep contributes to the wavefunction of states
in the ground state band is shown in Fig. 2 as a function of the angular
momentum of each state. Results obtained with the full basis are presented
in insert a), and with a basis truncated to 12 irreps in insert b). In
both cases the rotor parameters which fit the spectra best were used.
It is remarkable that the wavefunctions in the truncated basis are very
similar to those obtained using the full basis. This is an important
result because even for $^{20}$Ne, the textbook example of a `pure' SU(3)
rotor, the leading irrep (8,0) displays very large mixing  with the
irreps (6,1) with S=1 and, for J=8, with the irrep (4,2) with S=2. For J=6
the irrep (8,0) contributes only with 44 \%, and for J=8 with 23 \%.

The most important conclusion from the $^{20}$Ne results is that while
including a few SU(3) irreps in the basis is enough to obtain good
wavefunction, irreps with spin 1 and 2 and largest $C_2$ values must be
present. This particular selection of states is what we identify as the {\em
the quasi SU(3) truncation scheme}. This appears to be at work in all the
nuclei studied in the present work.

\section{The $^{22}${Ne} case}

The $^{22}${Ne} nucleus has 2 protons and 4 neutrons occupying the sd shell.
The proton SU(3) irreps are the same as for $^{20}${Ne}. The 4 neutrons can
be configured in any one of the 10 irreps listed in Table 4. By coupling the
proton and neutron irreps one gets a complete basis, which  contains 307
irreps, including the external multiplicity $\rho$ and all allowed spins.
Results are presented in Table 5 for a truncated basis with only the 13
irreps with largest $C_2$ values. Notice that only proton and neutron irreps
with spin 0 and 1 appear in the list (irreps with spin 2 have always a very
small $C_2$ value). These couple to irreps with spin 0, 1 and 2.

The $^{22}$Ne energy spectra is presented in Fig. 3. The first column
on the left hand side shows the energies obtained using the truncated
basis, the second the energies obtained using the complete basis, the
third the experimental energies \cite{NNDC} and the fourth the energies 
obtained using full $sd$-shell model calculations \cite{Pre72}. Insert 
a) displays the ground state and
$\beta$ bands, insert b) other excited bands. The ground state and the
$\beta$ bands are very well described with both the complete and the
truncated basis. Other states like $1^+_1$,  $3^+_1$, $4^+_2$, $3^+_2$,
$0^+_2$ and $4^+_3$ are also reasonable well predicted. In contrast,
for a number of highly excited states ($5^+_1$, $1^+_2$, $6^+_2$, $0^+_3$,
$0^+_4$) the model fails when the truncated basis is used. This is
particularly so for the states $2^+_4$, $1^+_2$, $0^+_3$ and $0^+_4$.
Nonetheless, the overall description of $^{22}$Ne is quite good.

\begin{table}
\begin{tabular}{cl}
S & irreps \\ \hline
0 & (4,2), (0,4), (3,1), (2,0)\\
1 & (5,0), (3,3), (3,1), (1,2), (0,1)\\
2 & (1,2)
\end{tabular}
\caption{SU(3) irreps for 4 particles in the sd-shell. }
\end{table}

\begin{table}
\begin{tabular}{ccccc|ccccc}
$(\lambda_\pi, \mu_\pi )$&$(\lambda_\nu, \mu_\nu )$&$(\lambda, \mu )$&
$S$ & $C_2$
&$(\lambda_\pi, \mu_\pi )$&$(\lambda_\nu, \mu_\nu )$&$(\lambda, \mu )$&
$S$ & $C_2$
  \\ \hline
(4,0) & (4,2) & (8,2) & 0   & 114       &
(4,0) & (5,0) & (9,0) & 1   & 108      \\
(4,0) & (4,2) & (6,3) & 0   & 90   &
(4,0) & (2,3) & (6,3) & 1   & 90   \\
(2,1) & (4,2) & (6,3) & 1   & 90   &
(4,0) & (4,2) & (7,1) & 0   & 81   \\
(4,0) & (5,0) & (7,1) & 1   & 81    &
(4,0) & (3,1) & (7,1) & 0   & 81    \\
(4,0) & (3,1) & (7,1) & 1   & 81    &
(2,1) & (4,2) & (7,1) & 1   & 81    \\
(2,1) & (5,0) & (7,1) & 0   & 81    &
(2,1) & (5,0) & (7,1) & 1   & 81    \\
(2,1) & (5,0) & (7,1) & 2   & 81
\end{tabular}
\caption{ List of irreps included in the truncated basis for $^{22}{Ne}$.
The proton
$(\lambda_\pi, \mu_\pi )$, neutron $(\lambda_\nu, \mu_\nu )$ and coupled irreps
$(\lambda,\mu)$ are listed in the first three columns, the values of
the total spin S in the fourth, and the $C_2$ value in the fifth column.}
\end{table}

As was seen for the $^{20}$Ne case, the single particle energies
together with the pairing and quadrupole-quadrupole interactions suffices
to reproduce the gross features of the energy spectrum. Small corrections are
introduced by the rotor-like terms ($K^2$ for $^{20}$Ne, $J^2$ for $^{22}$Ne),
which allow for a fine tuning of the energies. This is another very important
result. It reflects on the fact that the Hamiltonian parameters taken from
systematics are quite good, allowing a description comparable to the $sd$-shell
model. When coupled with the truncation scheme outlined
above, they constitute the main ingredients of a powerful and predictive
model, the quasi-SU(3) scheme.

\begin{figure}
\vspace{-2cm}
\epsfxsize=18cm
\centerline{\epsfbox{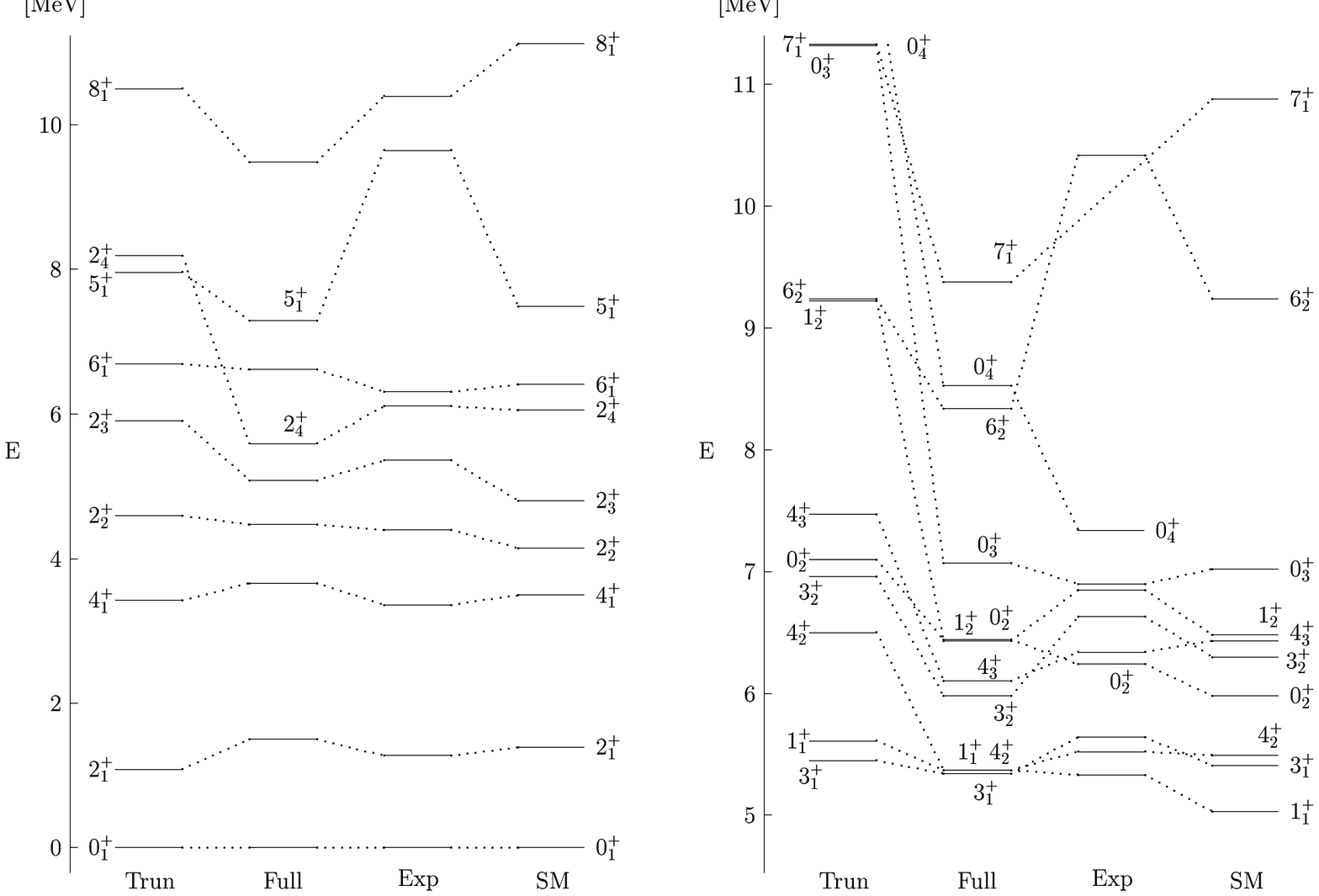}}\vspace*{-5.1cm}
\vspace{-7cm}
\caption{ Energy spectra of $^{22}$Ne. Insert a) shows the ground and
$\gamma$ bands, insert b) other excited bands.
Experimental energies are displayed in the third column, theoretical
results obtained using the truncated basis in the first column and
with the complete basis in the second column. Fourth column displays
the shell-model results \cite{Pre72}.}
\label{fig3}
\end{figure}

The $^{22}$Ne B(E2) transition strengths are presented in Table 6.
The theoretical results obtained with $q_{eff}=1.3$ agree well with the
known experimental data (there are three strengths measured between
states in the ground state band as well as two transitions between bands). As
was the case for
$^{20}$Ne, the B(E2) transition strengths between members of the ground
state band are pretty well described within both the truncated and the
complete Hilbert space, while the transitions between states in gamma
bands differ, with the complete space predictions being smaller.

\begin{table}
\begin{tabular}{c|ccc}
$J \rightarrow (J+2)$ &      Experimental    & Truncated & Complete\\ \hline
$0_1 \rightarrow 2_1$ & $2.2886 \pm 0.0915$  & 2.3192    &  2.0925    \\
$2_1 \rightarrow 4_1$ & $1.1650 \pm 0.0265$  & 1.0684    &  0.9419    \\
$4_1 \rightarrow 6_1$ & $0.7245 \pm 0.0898$  & 0.7628    &  0.6402    \\
$6_1 \rightarrow 8_1$ &                      & 0.4931    &  0.2885    \\
$0_1 \rightarrow 2_2$ & $>0.0476$            & 0.0241    &  0.0222    \\
$2_1 \rightarrow 4_2$ & $0.0062 \pm 0.0015$  & 0.0008    &  0.0024    \\
$2_2 \rightarrow 4_2$ &                      & 0.4489    &  0.0971    \\
$4_2 \rightarrow 6_2$ &                      & 0.5624    &  0.2073    \\
$6_2 \rightarrow 8_2$ &                      & 0.3639    &  0.1656    \\
\multicolumn{3}{c}{}\\
\end{tabular}
\caption{B(E2;J $\rightarrow$ (J+2)) [$e^2b^2 \times 10^{-2}$] transition
strengths for $^{22}{Ne}$. Experimental and theoretical values, both for
the truncated and the complete basis, are shown in columns 2, 3 and 4
respectively.}
\end{table}

The ground state wavefunctions obtained using the complete and
truncated bases are displayed in Fig. 4 as a function of the angular
momentum of the states. While the details may differ, a comparison of the
wavefunctions exhibit the same trends in both the complete and truncated
basis. For J=0 the irrep $(4,0)_\pi \otimes (4,2)_{\nu} \rightarrow (8,2)$
S=0 dominates, with significant mixing with the irreps (9,0) S=1, (6,3) S=1
and (7,1) S=1. For larger values of the angular momentum, the contribution of
the irrep (8,2) S=0 decreases, being replaced by the irrep (7,1) S=2 for
J = 8 and 10. The important role played by the SU(3) irreps with spin S = 1
and 2 in the ground state band is the most significant departure from the
usual Elliot SU(3) symmetry and, as mentioned above, is the manifestation of
the quasi-SU(3) symmetry.

\begin{figure}
\vspace*{-1cm}\hspace*{-0.5cm}
\epsfxsize=13cm
\centerline{\epsfbox{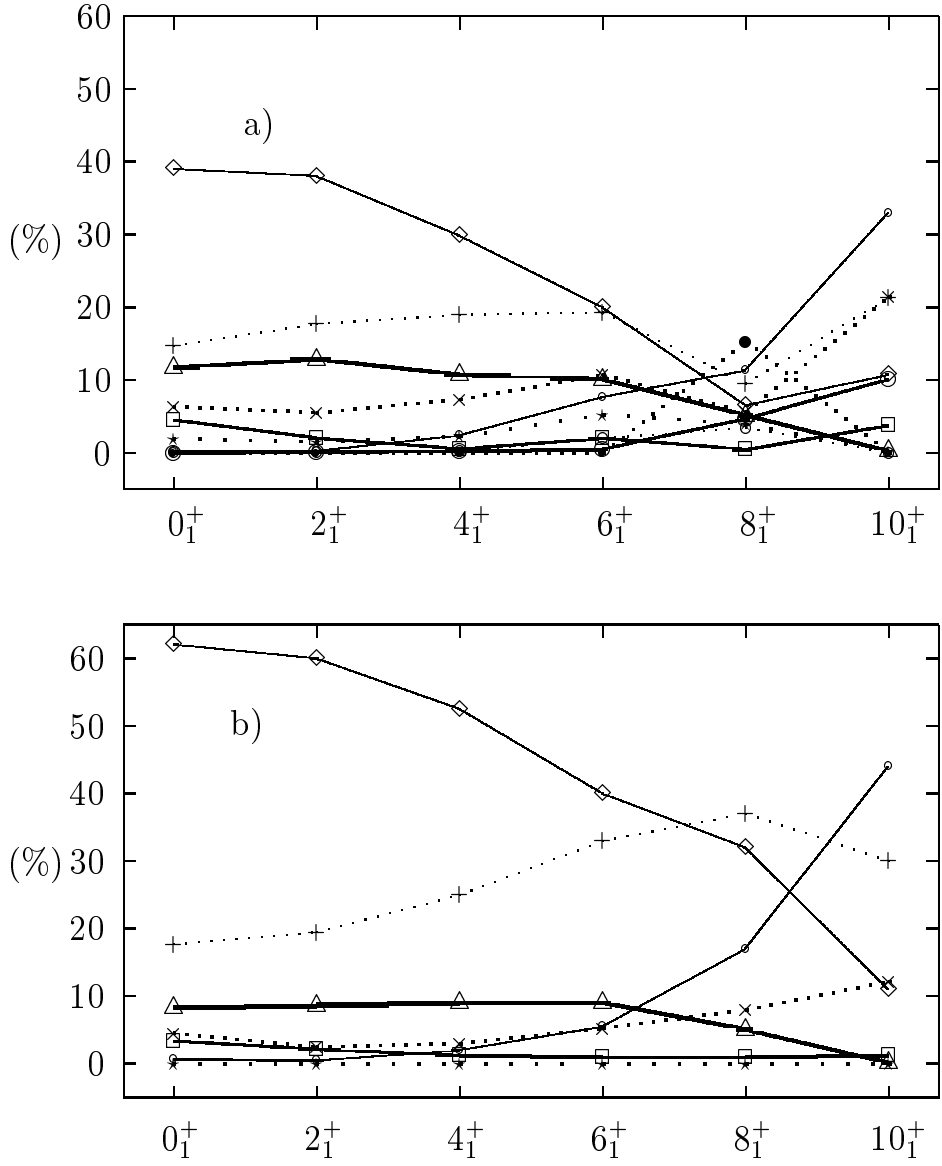}}\vspace*{-2cm}
\caption{Wavefunction components of states belonging to the ground state
band in $^{22}$Ne. The percentage each irrep contributes to the wavefunction
is shown as function of the angular momentum for the complete basis, insert
a)  and the truncated basis, insert b). The convention used is: $\diamond$ for
(8,2)0;
$+$ for (9,0)1; $\sqcap \!\!\!\! \sqcup$ for [(4,0)$\otimes$ (2,3)] (6,3)1;
$\times$ for [(2,1)$\otimes$ (4,2)] (6,3)1;
$\triangle$ for [(4,0) $\otimes$ (3,1)] (7,1)1;
$\star$ for [(2,1) $\otimes$ (4,2)] (7,1)1;
$\circ$ for (7,1)2; $\odot$ for (6,0)2;
$\bigcirc$ for (4,4)2; and  $\bullet$ for (5,2)2, (3,3)2,3.}
\end{figure}

\section{The $^{24}${Mg} case}

The $^{24}${Mg} nucleus has four protons and four neutrons in the active
sd shell. The 10 SU(3) irreps for protons and neutrons are those listed in
Table 4. The complete basis includes 1599 irreps, taking into account the
external multiplicity $\rho$ and the different spin values. Calculations were
carried out in truncated Hilbert space built with the 11 SU(3) irreps listed
in Table 7. Again, these are the SU(3) irreps with the largest $C_2$ value.
Proton and neutron irreps either spin S=0 and 1 and the coupled irreps have
spin S = 0, 1 and 2.

\begin{table}
\begin{tabular}{ccccc|ccccc}
$(\lambda_\pi, \mu_\pi )$&$(\lambda_\nu, \mu_\nu )$&$(\lambda, \mu )$&
$S$ & $C_2$
&$(\lambda_\pi, \mu_\pi )$&$(\lambda_\nu, \mu_\nu )$&$(\lambda, \mu )$&
$S$ & $C_2$
  \\ \hline
(4,2) & (4,2) & (8,4) & 0   & 148       &
(4,2) & (4,2) & (9,2) & 0   & 136      \\
(4,2) & (5,0) & (9,2) & 1   & 136      &
(5,0) & (4,2) & (9,2) & 1   & 136      \\
(4,2) & (4,2) & (10,0) & 0   & 130      &
(5,0) & (5,0) & (10,0) & 0   & 130      \\
(5,0) & (5,0) & (10,0) & 1   & 130      &
(5,0) & (5,0) & (10,0) & 2   & 130      \\
(4,2) & (4,2) & (6,5) & 0   & 124      &
(4,2) & (2,3) & (6,5) & 1   & 124      \\
(2,3) & (4,2) & (6,5) & 1   & 124
\end{tabular}
\caption{ List of irreps included in the truncated basis for $^{24}{Mg}$.
The proton
$(\lambda_\pi, \mu_\pi )$, neutron $(\lambda_\nu, \mu_\nu )$ and coupled irreps
$(\lambda,\mu)$ are listed in the first three columns, the values of
the total spin S in the fourth, and the $C_2$ value in the fifth column.}
\end{table}

As can be seen in Table 2, the Hamiltonian parameters include an symmetry
parameter. This is required to lower the energy of the $8^+_1$ state, which
is a member of the ground state band. The energy spectra of $^{24}$Mg is
shown in Fig. 5. The first column displays the shell-model values, the
second the experimental data, and the third the predicted energies in the
truncated basis. The last three columns show the effect on the spectra of
turning off the $K^2$ term ($a=0$), both the $K^2$ and the symmetry term
($a=A_{sym} = 0$) and the three rotor terms ($a = b = A_{sym} = 0$),
respectively.

\begin{figure}\vspace{-1cm}
\epsfxsize=12.3cm
\centerline{\epsfbox{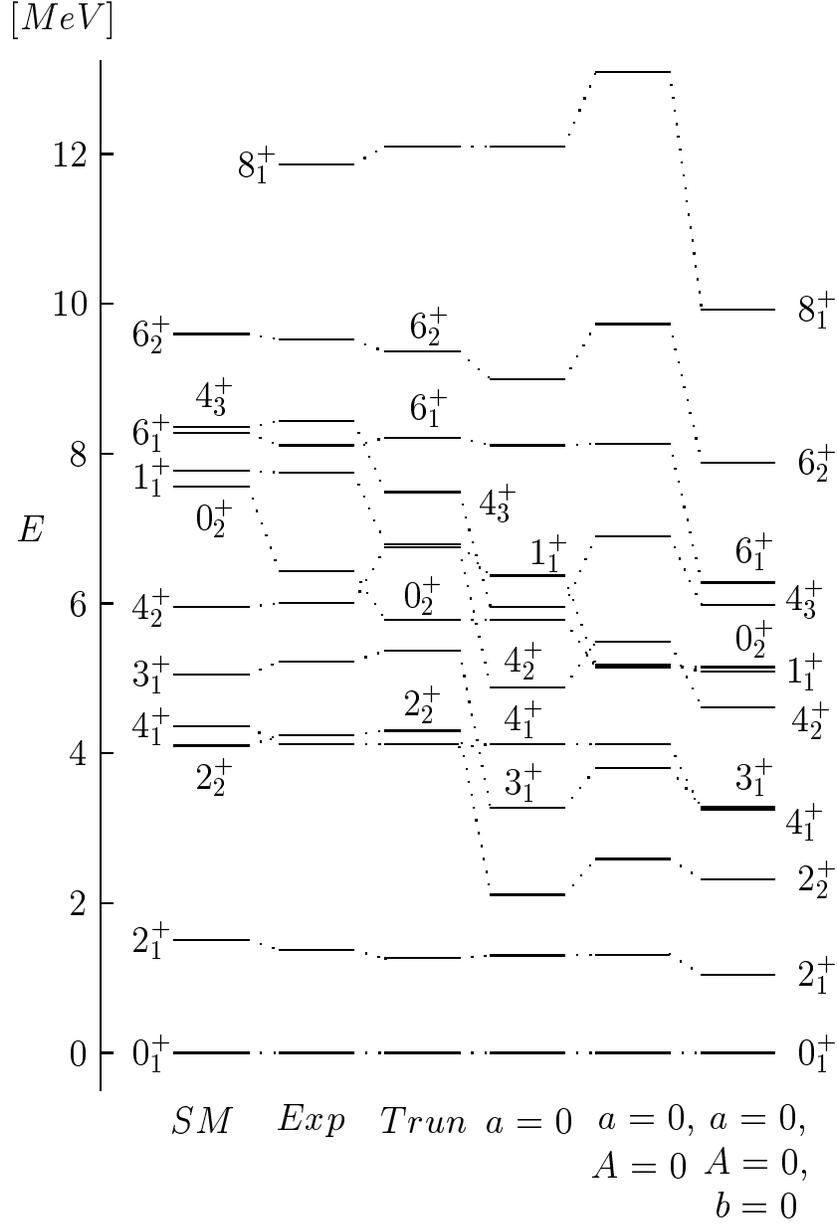}}\vspace{-0.8cm}
\caption{Energy spectra of $^{24}$Mg. The first column displays the
full shell-model values, the second one the
experimental data, the third the predicted energies in the truncated
basis. The last three columns show the effect on the energy spectra of
turning off the $K^2$ term ($a=0$), both the $K^2$ and the symmetry term
($a=A_{sym} = 0$) and the three rotor terms ($a = b = A_{sym} = 0$),
respectively. The $SM$ values were taken from \cite{Wil84,End90}.}
\end{figure}

\begin{figure}
\vspace{-2cm}
\epsfxsize=13.4cm
\centerline{\epsfbox{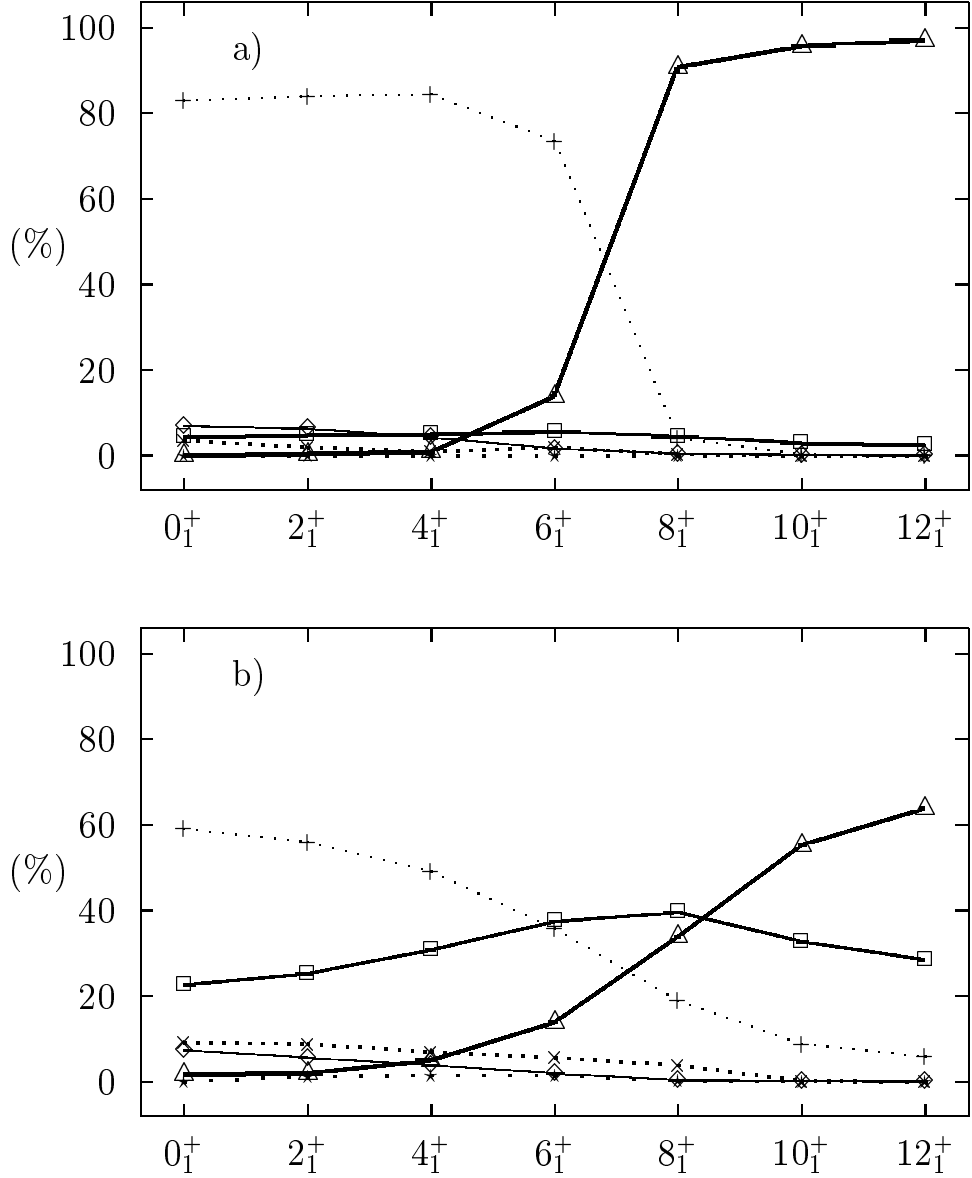}}\vspace{-2cm} 
\caption{Wave function composition of states in the ground state band of
$^{24}$Mg as a function of their angular momentum. Insert a) shows the
results using the complete Hamiltonian (\ref{eq:ham}), insert b) the
results setting  $a = b = A_{sym} = 0$. The percentage each irrep
contributes is shown on the vertical axis. The convention used is
$\diamond$ for (10,0)0; $+$ for (8,4)0; $\sqcap \!\!\!\! \sqcup$ for (9,2)1;
$\times$ for (6,5)1; $\triangle$ for (10,0)2; and $\star$ for (10,0)0.}
\end{figure}

\begin{table}
\begin{tabular}{c|cc}
$J \rightarrow (J+2)$           & Experimental  & Theoretical \\ \hline
$0^+_1~ \rightarrow ~2^+_1$&~~~~$4.3588 \pm 0.1028$~~~~&4.7742  \\
$2^+_1~ \rightarrow ~4^+_1$     & $2.6915 \pm 0.3738$ & 2.3138  \\
$4^+_1~ \rightarrow ~6^+_1$     & $2.3165 \pm 0.8316$ & 1.7052  \\
$6^+_1~ \rightarrow ~8^+_1$     &                     & 0.5366  \\
$8^+_1~ \rightarrow ~10^+_1$    &                     & 1.1093  \\
$10^+_1~ \rightarrow ~12^+_1$   &                     & 0.6164  \\
$0^+_1~ \rightarrow ~2^+_2$     & $0.3310 \pm 0.0226$ & 0.5022  \\
$0^+_1~ \rightarrow ~2^+_3$     & $0.1172 \pm 0.0514$ & 0.0004  \\
$0^+_2~ \rightarrow ~2^+_1$     & $0.0214 \pm 0.0033$ & 0.0005  \\
$0^+_2~ \rightarrow ~2^+_2$     & $0.2919 \pm 0.0493$ & 0.0590  \\
$0^+_2~ \rightarrow ~2^+_3$     &                    Ý& 3.2228  \\
$0^+_2~ \rightarrow ~2^+_4$     & $2.6729 \pm 1.0280$ & 1.3583  \\
$2^+_1~ \rightarrow ~4^+_2$     & $0.0733 \pm 0.0052$ & 0.0531  \\
$2^+_1~ \rightarrow ~4^+_3$     & $0.0747 \pm 0.0299$ & 0.0075  \\
$2^+_1~ \rightarrow ~4^+_4$     & $0.0254 \pm 0.0075$ & 0.0045  \\
$2^+_2~ \rightarrow ~4^+_2$     & $0.9495 \pm 0.0822$ & 0.8766  \\
$2^+_2~ \rightarrow ~4^+_3$     & $0.0972 \pm 0.0448$ & 0.0067  \\
$2^+_2~ \rightarrow ~4^+_4$     & $0.0067 \pm 0.0029$ & 0.0007  \\
$2^+_3~ \rightarrow ~4^+_3$     & $2.9159 \pm 1.1962$ & 1.5760  \\
$2^+_3~ \rightarrow ~4^+_4$     & $2.1681 \pm 0.6729$ & 0.6478  \\
$4^+_1~ \rightarrow ~6^+_2$     & $0.0356 \pm 0.0178$ & 0.1167  \\
$4^+_2~ \rightarrow ~6^+_2$     & $1.0691 \pm 0.5346$ & 0.0058  \\
$6^+_1~ \rightarrow ~8^+_2$     &                    & 0.8532  \\
$6^+_2~ \rightarrow ~8^+_1$     &                    & 1.0034  \\
$6^+_2~ \rightarrow ~8^+_2$     &                    & 0.4719
\end{tabular}
\caption{B(E2) transition strengths for $^{24}{Mg}$ (in [$e^2b^2
\times 10^{-2}$]). $q_{eff}$ is 1.6.}
\end{table}

The present theoretical spectra (`Trun') reproduce the observed energy
levels and the full shell-model calculations fairly well.
As mentioned above, gross features of the spectra are reproduced by the
dominant terms in Hamiltonian (\ref{eq:ham}): the single particle, pairing
and quadrupole-quadrupole interactions, whose strength are fixed from
systematics. The spectra obtained with no rotor-like terms is shown in the
last column (right hand side) of Fig. 5. The ordering of the levels is
correct, with only one crossing of two neighboring levels: $1^+_1$ and
$4^+_2$. The first $2^+$ state is about at the right energy. However,
without the rotor-like terms the model fails to reproduce the moment of
inertia of the ground state band, the energy of the $\gamma$ band head and
other features. The energy of the state in the $\gamma$ band is substantially
controlled by the $K^2$ term. This is why it was introduced \cite{Naq90}.
Without only this term the $\gamma$ band energies are clearly depressed, as
seen in the third column. The most prominent effect of not including the
symmetry term is in the energy of the $8^+_1$ state. This term reduces the
mixing of the leading irrep (8,4) S = 1 with other irreps which have
$\lambda$ or $\mu$ odd, a mixing that is driven by the single particle
terms of the Hamiltonian. It was also noted in a study of the mapping of the
rotor to the SU(3) model \cite{Les87}. It is shown here to contribute to the
correct prediction of the energies of high energy states belonging to the
ground state band.

Fig. 6 shows the percentage each irrep contributes to the wavefunction
of states in the ground state band of $^{24}$Mg as a function of their
angular momentum. The upper panel show the results using the complete
Hamiltonian (\ref{eq:ham}), the lower panel the results of setting  $a = b =
A_{sym} = 0$. The introduction of the symmetry term strongly diminishes
the mixing. There is a sudden change in the wavefunction from J=6 to J=8,
where the dominance of the (8,4) S=0 irrep is replaced by the (10,0) S=2.
This change in the spin contribution to states in the ground state band was
also found in full shell-model calculations of $^{24}$Mg \cite{Joh00}.

Calculations were performed also for other basis sets, containing 1, 4, 8,
26, 34, 54, and 62 irreps. In all cases similar fits were found, with the
parameter $b$ ranging between 0.01 and 0.03 and $A_{sym}$  between 0.04 and
0.08.

The B(E2) transition strengths for $^{24}{Mg}$ are listed in Table 8.
Experimental values are shown in the second column. Theoretical values were
calculated using an effective charge $q_{eff} = 1.6$ and are displayed in the
third column. The overall agreement is reasonable. The change in the
wavefunction of the $8^+_1$ state reflects in a fragmentation of the B(E2)
strengths between the transitions $6^+_1 \rightarrow 8^+_1$ and $6^+_1
\rightarrow 8^+_2$. None of these have been measured.

\section{The $^{28}${Si} case}

The $^{28}$Si nucleus has 6 protons and 6 neutrons occupying the sd shell.
Six like particles in the $sd$-shell populate the 15 irreps listed  with
their spin S in Table 9.

\begin{table}
\begin{tabular}{cl}
S & irreps \\ \hline
0 & (6,0), (0,6), (3,3), (2,2), (0,0)\\
1 & (3,3), (4,1), (1,4), (2,2), (3,0), (0,3), (1,1)\\
2 & (2,2), (1,1) \\
3 & (0,0)
\end{tabular}
\caption{SU(3) irreps for 6 particles in the sd shell. }
\end{table}

The complete basis includes 4045 irreps, taking into account the
external multiplicity $\rho$ and the different spin values. Calculations were
carried out in a severely truncated Hilbert space, built with the 24 SU(3)
irreps listed in Table 10. Again, these are the SU(3) irreps with the largest
$C_2$ values.  As in previous cases, proton and neutron irreps include those
with spin S = 0 and 1, while total irreps have spin S = 0, 1 and 2. The proton
and neutron spin $S_\pi, S_\nu$ are shown explicitly in order to distinguish
the irreps (3,3) with spin S = 0 and 1.

\begin{table}
\begin{tabular}{ccccc|ccccc}
$(\lambda_\pi, \mu_\pi )S_\pi$&$(\lambda_\nu, \mu_\nu )S_\nu$&$(\lambda,
\mu )$&
$S$ & $C_2$
&$(\lambda_\pi, \mu_\pi )S_\pi$&$(\lambda_\nu, \mu_\nu )S_\nu$&$(\lambda,
\mu )$&
$S$ & $C_2$
  \\ \hline
(6,0)0 & (6,0)0 & (12,0) & 0   & 180       &
(0,6)0 & (0,6)0 & (0,12) & 0   & 180      \\
(6,0)0 & (3,3)0 & (9,3) & 0   & 153   &
(6,0)0 & (3,3)1 & (9,3) & 1   & 153  \\
(0,6)0 & (3,3)0 & (3,9) & 0   & 153  &
(0,6)0 & (3,3)1 & (3,9) & 1   & 153   \\
(3,3)0 & (6,0)0 & (9,3) & 0   & 153   &
(3,3)1 & (6,0)0 & (9,3) & 1   & 153   \\
(3,3)0 & (0,6)0 & (3,9) & 0   & 153   &
(3,3)1 & (0,6)0 & (3,9) & 1   & 153   \\
(6,0)0 & (6,0)0 & (10,1) & 0  & 144   &
(6,0)0 & (0,6)0 & (6,6) & 0  & 144   \\
(0,6)0 & (6,0)0 & (6,6) & 0  & 144   &
(0,6)0 & (0,6)0 & (1,10) & 0  & 144   \\
(6,0)0 & (4,1)1 & (10,1) & 1  & 144   &
(4,1)1 & (6,0)0 & (10,1) & 1  & 144   \\
(0,6)0 & (1,4)1 & (1,10) & 1  & 144   &
(1,4)1 & (0,6)0 & (1,10) & 1  & 144   \\
(3,3)0 & (3,3)0 & (6,6) & 0  & 144   &
(3,3)1 & (3,3)1 & (6,6) & 0  & 144   \\
(3,3)1 & (3,3)1 & (6,6) & 1  & 144   &
(3,3)1 & (3,3)1 & (6,6) & 2  & 144   \\
(3,3)0 & (3,3)1 & (6,6) & 1  & 144   &
(3,3)1 & (3,3)0 & (6,6) & 1  & 144
\end{tabular}
\caption{ List of irreps included in the truncated basis for $^{28}{Si}$.
The proton  $(\lambda_\pi, \mu_\pi )$ and neutron $(\lambda_\nu, \mu_\nu )$
irreps with their spins $S_\pi$ and $S_\nu$, and coupled irreps
$(\lambda,\mu)$ are listed in the first three columns, the values of
the total spin S in the fourth, and the $C_2$ value in the fifth column.}
\end{table}

\begin{figure}
\vspace{-1cm}
\epsfxsize=17cm
\centerline{\epsfbox{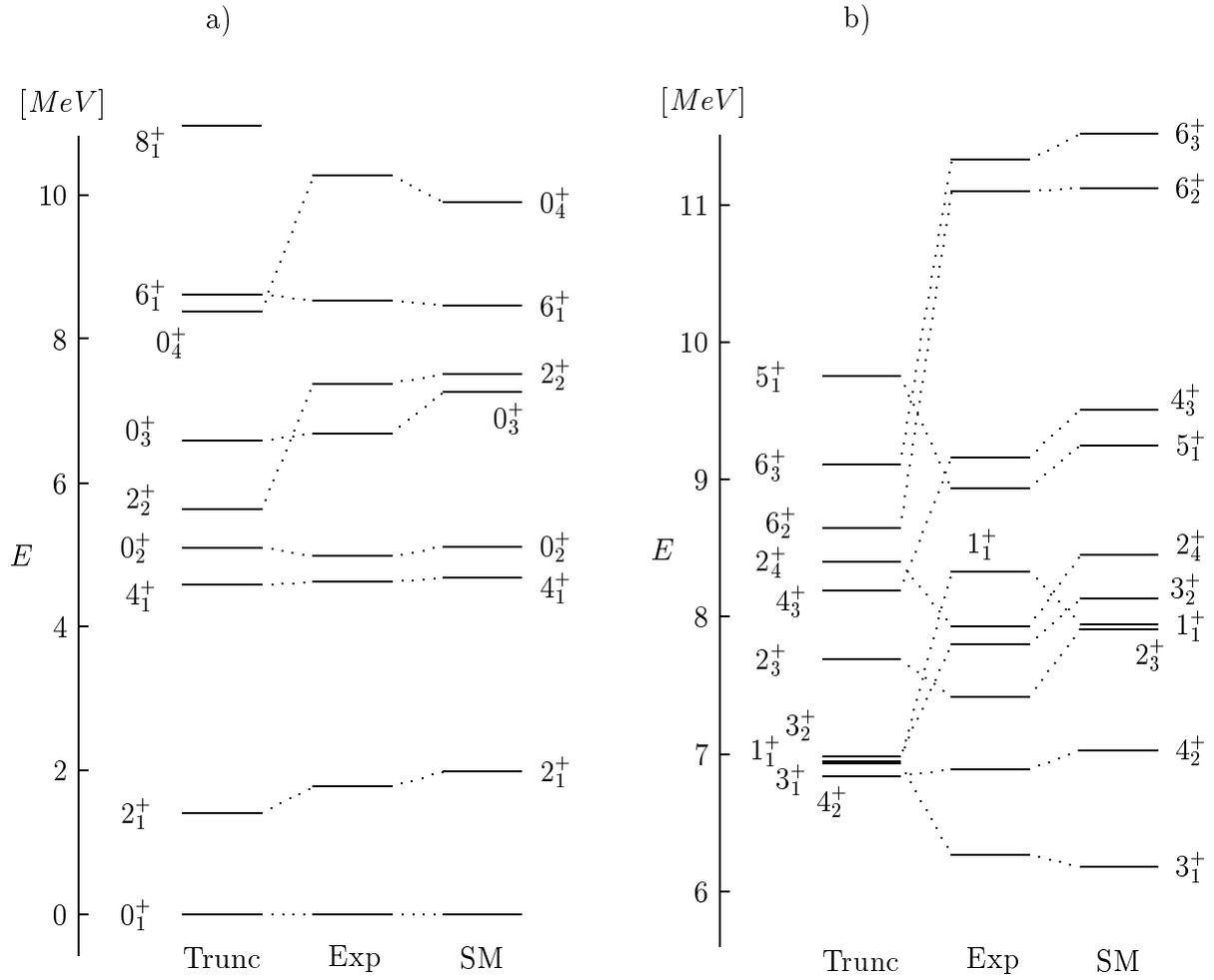}}\vspace*{-8.6cm}
\caption{Energy spectra for $^{28}$Si. Experimental results are shown in
the first column, theoretical energies calculated with the present model
are displayed in the second column, and those obtained with a full shell
model calculation \cite{Wil84,End90} in the third.}
\label{fig7}
\end{figure}

\begin{figure}
\vspace*{-1cm}\hspace*{-0.5cm}
\epsfxsize=14cm
\centerline{\epsfbox{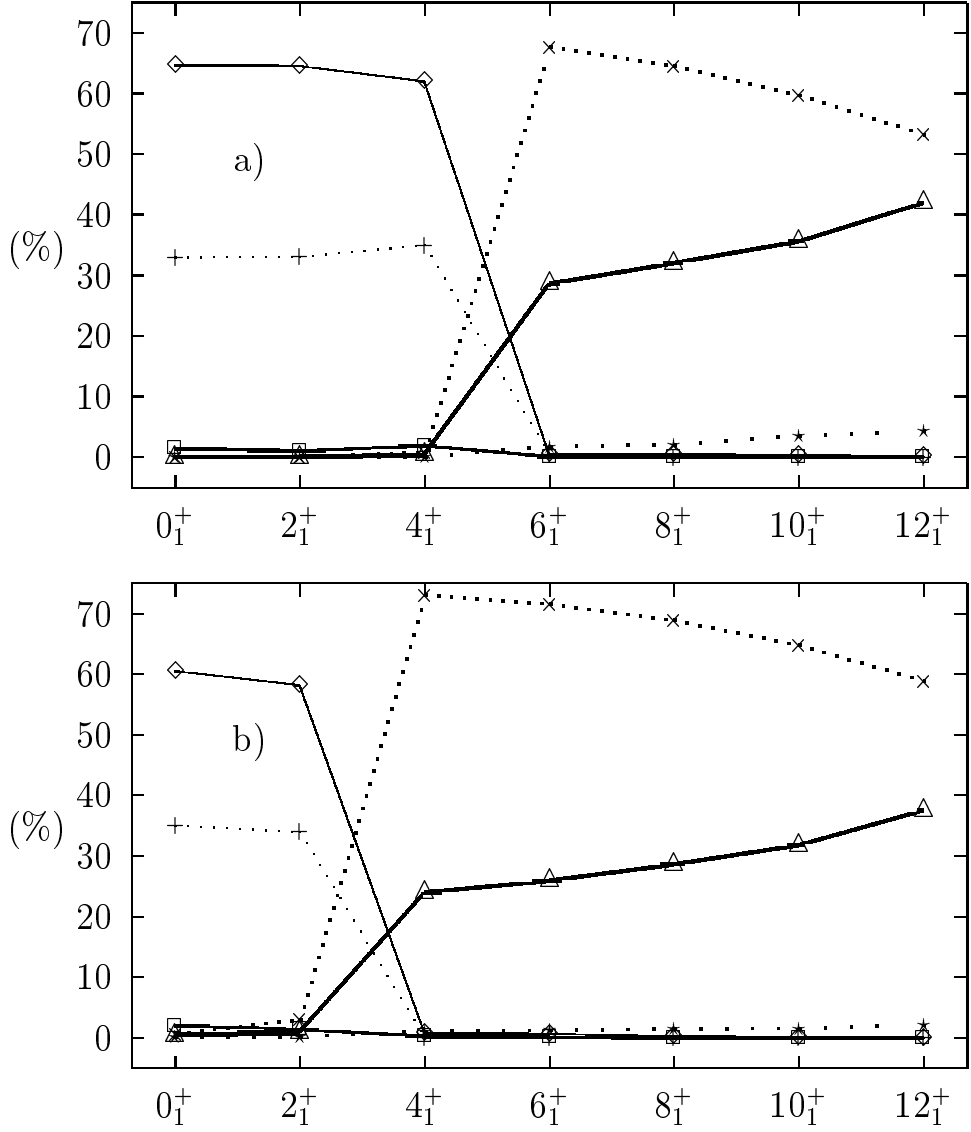}}\vspace*{-3cm}
\caption{Wavefunction components of states belonging to the ground state
band in $^{28}$Si. The percentage each irrep contributes is shown as a
function of the angular momentum for the basis with 24 irreps, in insert a)
with the Hamiltonian shown in Table \ref{t1} plus the $C_3$ term and
insert b) for the same
Hamiltonian but without $C_3$.  The convention used is $\diamond$ for
(0,12)0; $+$ for (1,10)1; $\sqcap \!\!\!\! \sqcup$ for
(12,0)0, $\times$ for (10,1)0, $\triangle$ for (2,8)2, and $\star$ for
(8,2)2.}
\label{fig8}
\end{figure}

\begin{table}
\begin{tabular}{c|cc}
$J \rightarrow (J+2)$           & Experimental  & Theoretical \\ \hline
$0^+_1~ \rightarrow ~2^+_1$ &~~~$3.3334 \pm 0.0757$~~~& 4.9208  \\
$2^+_1~ \rightarrow ~4^+_1$     & $1.2671 \pm 0.1193$ & 2.3668  \\
$4^+_1~ \rightarrow ~6^+_1$     & $0.7222 \pm 0.1824$ & 0.0004  \\
$0^+_1~ \rightarrow ~2^+_2$     & $0.0780 \pm 0.0150$ & 0.0087  \\
$0^+_1~ \rightarrow ~2^+_3$     & $0.0445 \pm 0.0055$ & 0.0393  \\
$0^+_1~ \rightarrow ~2^+_4$     & $0.0755 \pm 0.0075$ & 0.0021  \\
$0^+_1~ \rightarrow ~2^+_5$     & $0.0073 \pm 0.0020$ & 0.0026  \\
$0^+_1~ \rightarrow ~2^+_6$     & $0.0130 \pm 0.0040$ & 0.0004  \\
$0^+_1~ \rightarrow ~2^+_7$     & $0.0355 \pm 0.0150$ & 0.0082  \\
$0^+_2~ \rightarrow ~2^+_1$     & $0.4343 \pm 0.0808$ & 0.0006  \\
$0^+_2~ \rightarrow ~2^+_2$     & $0.1765 \pm 0.0760$ & 4.9371  \\
$0^+_2~ \rightarrow ~2^+_4$     & $0.5050 \pm 0.5810$ & 0.0001  \\
$0^+_2~ \rightarrow ~2^+_5$     & $1.3890 \pm 0.3280$ & 0.0016  \\
$0^+_2~ \rightarrow ~2^+_7$     & $0.0757 \pm 0.0430$ & 0.0001  \\
$0^+_2~ \rightarrow ~2^+_9$     & $0.0255 \pm 0.0125$ & 0.0001  \\
$0^+_3~ \rightarrow ~2^+_1$     & $0.0227 \pm 0.0035$ & 0.0042  \\
$0^+_4~ \rightarrow ~2^+_1$     & $ > 0.0019        $ & 0.0025  \\
$2^+_1~ \rightarrow ~4^+_4$     & $0.0007 \pm 0.0001$ & 0.0031   \\
$2^+_2~ \rightarrow ~4^+_4$     & $0.0202 \pm 0.0045$ & 0.0004  \\
$2^+_3~ \rightarrow ~4^+_3$     & $0.9825 \pm 0.1653$ & 0.1358  \\
$2^+_4~ \rightarrow ~4^+_1$     & $0.1212 \pm 0.0138$ & 0.0001  \\
$2^+_4~ \rightarrow ~4^+_4$     & $2.2037 \pm 0.4591$ & 0.0049  \\
$2^+_5~ \rightarrow ~4^+_1$     & $0.0734 \pm 0.0274$ & 0.0013  \\
$2^+_7~ \rightarrow ~4^+_1$     & $0.0146 \pm 0.0066$ & 0.0002  \\
$2^+_9~ \rightarrow ~4^+_1$     & $0.0197 \pm 0.0081$ & 0.0001  \\
$2^+_4~ \rightarrow ~4^+_6$     & $0.1212 \pm 0.0252$ & 0.1705  \\
\end{tabular}
\caption{B(E2) transition strengths for $^{28}{Si}$ ([$e^2b^2
\times 10^{-2}$] with $q_{eff}$ is 1.3.)}
\label{tab11}
\end{table}

The $^{28}$Si energy spectra is shown in Fig. 7. In was obtained with
the parametrization given in Table \ref{t1} for the Hamiltonian
(\ref{eq:ham}) plus a term proportional to $C_3$, the third order SU(3)
Casimir operator. The $C_3$ eigenvalues go like cubic powers of $\lambda$
and $\mu$ \cite{Tro95}, being in general large numbers. For this reason
its coefficient was selected to be very small: 4.64 $\times 10^{-4}$.
Its effect on the wavefunction can be appreciated in Figure \ref{fig8}.
It drives irreps with $\mu \gg \lambda$ lower in energy than those with $\mu
\ll \lambda$. As this is a mid-shell nucleus so it has the same number of
particles and holes, it has two bands originated from the (12,0) and (0,12)
irreps which would be degenerated for a pure quadrupole-quadrupole interaction.
A similar feature has been found in Hartree-Fock calculations, where two minima
coexist \cite{Das67}. It was shown that the $l \cdot s$ single particle term
breaks this degeneracy and favors the dominance in the ground state band of
the (0,12) irrep \cite{Das67}. But the $l \cdot s$ term alone is unable to push
the first excited $0^+$ state to an energy of 4.98 MeV. We found that, at
most, it breaks the degeneracy putting this state at 1.16 MeV. A similar trend
is observed in other states belonging to the first excited (prolate) band. 
The term proportional to $C_3$ allows us to put the first excited $0^+$
at the correct energy, as can be seen in Figure \ref{fig7}.
With this Hamiltonian we obtain a very good description of this nucleus.
Some states ($1_1$, $2_3$, $2_4$ and $5_1$) deviate from the
experimental values in a way that is similar to those predicted by
shell-model calculations \cite{Wil84,End90}.

In Figure \ref{fig8} the crossing of the oblate [(0,12),(1,10)] and prolate
[((12,0), (10,1)] bands can be seen, which is a special feature of this
mid-shell nucleus. With the $C_3$ term in the Hamiltonian, the crossing in
Figure \ref{fig8} insert a) occurs at J = 6, while without this term, insert
b), it occurs at J = 4. This happen because $C_3$ favors the J = 4 state
belonging to the (0,12) band, driving it lower in energy than its
counterpart from (12,0). Without $C_3$ there is a state  with J = 4 at 4.58 MeV
dominated by the (0,12) irrep, as for the calculation with $C_3$, but in this
case there is another J = 4 state dominated by the (12,0) irrep at 2.87 MeV.

As it was the case for lighter nuclei, the energy spectra is reproduced fairly
well, and the quality of the results is comparable with those obtained by
Wildenthal and Endt \cite{Wil84,End90} with a full shell-model 
calculation.
The ground state band and many excited states are properly described. Given
that this nucleus lies exactly at mid-shell, it is the  most complex
even-even nuclei of the sd shell. The presence of a $3^+_1$ state that is
lower in energy than the $2^+_2$ state, i.e. which does not belong to the
$\gamma$ band, is impossible to understand within the context of the present
model, while shell-model calculations reproduce this state. Similar
problems occur for other excited states, like the $2^+_3, 6^+_2, 6^+_3$.

In Table \ref{tab11} the B(E2) transition strengths are shown. While there are
many measured transitions, mostly of them involve the $0^+$, $2^+$ and $4^+$
states.  The agreement between theory and experiment is not as good as in the
other nuclei, but shows the same trend found in full shell-model 
calculations for $^{28}$Si. There are few B(E2) experimental intensities
with large strength, they allow us to identify some bands
like those formed by the set of states ($0^+_1$, $2^+_1$, $4^+_1$,
$6^+_1$); ($2^+_2$, $4^+_3$) and ($2^+_4$, $4^+_4$). On other hand,
theoretical strengths suggest the following associations between positive
parity  states: ($0_1$, $2_1$, $4_1$, $6_3$, $8_5$); ($0_2$, $2_2$, $4_2$,
$6_2$, $8_1$, $10_1$); ($0_3$, $2_3$, $4_6$, $6_7$, $8_{11}$, $10_6$);
($0_4$, $2_6$, ($4_7$, $4_9$)), where a couple of states with J = 4 in
parentheses means that there are large strengths toward both states.

\section{Conclusions}

The quasi SU(3) symmetry concept, first introduced in full shell-model
calculations, was used in this work to explore the structure of four
even-even sd-shell nuclei. While the SU(3) symmetry of these nuclei has been
used since 1959, the new feature in this work is a close look at the role
of proton and neutron irreps with spin S=1. It turns out that these
configurations make large, often dominant contributions to the structure
of members of the ground-state band, especially those states with higher
angular momentum values.

The energy spectra and B(E2) transition strengths of $^{20,22}$Ne, $^{24}$Mg
and $^{28}$Si were calculated in a truncated basis based on the quasi
SU(3) symmetry. A simple Hamiltonian which contains realistic
single-particle energies, pairing and quadrupole-quadrupole terms was
used. Both, the energy spectra and B(E2) transition were
found to be very close to the corresponding experimental values. For the
$^{20,22}$Ne cases the calculations were performed in the complete as well as
a quasi SU(3) truncated basis, allowing for a discussion of the benefits and
limitations of the truncation scheme. While including a few SU(3) irreps
in the basis is enough to obtain good description of deformed nuclei,
irreps with spin 1 and 2 and largest $C_2$ values must be present.
Effects of the three rotor-like terms in the Hamiltonian were discussed
in detail for the $^{24}$Mg case.

The results of this work re-enforce the claim that the quasi SU(3) symmetry
is a very powerful concept, raising expectations for its use in a description
of intruder states in heavy deformed nuclei. Together with a pseudo SU(3)
description for nucleons in normal parity states in these nuclei, one can
anticipate a relatively simple SU(3) description (pseudo for normal parity
orbitals and quasi for intruders) of heavy deformed nuclei which has been
missing up to now.

\section{Aknowledgements}

This work was supported in part by Conacyt (M\'exico) and the National
Science Fundation under Grant PHY-9970769 and Cooperative Agreement 
EPS-9720652 that includes matching from the Louisiana Board of Regents 
Support Fund.

\bigskip

\end{document}